# Quantum support vector machines for classification and regression on a trapped-ion quantum computer

Teppei Suzuki[1] · Takashi Hasebe[1] · Tsubasa Miyazaki[1]

Abstract

Quantum machine learning is a rapidly growing field at the intersection of quantum computing and machine learning. In this work, we examine our quantum machine learning models, which are based on quantum support vector classification (QSVC) and quantum support vector regression (QSVR). We investigate these models using a quantum-circuit simulator, both with and without noise, as well as the IonQ Harmony quantum processor. For the QSVC tasks, we use a dataset containing fraudulent credit card transactions and image datasets (the MNIST and the Fashion-MNIST datasets); for the QSVR tasks, we use a financial dataset and a materials dataset. For the classification tasks, the performance of our QSVC models using 4 qubits of the trapped-ion quantum computer was comparable to that obtained from noiseless quantum-circuit simulations. The result is consistent with the analysis of our device-noise simulations with varying qubit-gate error rates. For the regression tasks, applying a low-rank approximation to the noisy quantum kernel, in combination with hyperparameter tuning in $\varepsilon$-SVR, improved the performance of the QSVR models on the near-term quantum device. The alignment, as measured by the Frobenius inner product between the noiseless and noisy quantum kernels, can serve as an indicator of the relative prediction performance on noisy quantum devices in comparison with their ideal counterparts. Our results suggest that the quantum kernel, as described by our shallow quantum circuit, can be effectively used for both QSVC and QSVR tasks, indicating its resistance to noise and its adaptability to various datasets.

To whom correspondence should be addressed:

Teppei Suzuki

`tep.suzuki@scsk.jp`

[1] *Technology Strategy Division, SCSK Corporation, Toyosu Front, 3-2-20 Toyosu, Koto-ku, Tokyo 135-8110, Japan*

# 1 Introduction

The field of quantum technologies (Nielsen and Chuang 2010) has seen tremendous progress in recent years, with the potential to transform a wide range of scientific research and industries. One possible application of quantum computing is the field of quantum machine learning (Rebentrost et al. 2014; Biamonte et al. 2017; Cerezo et al. 2022), which could potentially be used for classifying and recognizing complex patterns more efficiently than classical methods (Abbas et al. 2021). The quantum kernel method, a candidate in the field, leverages quantum states as described by quantum circuits to compute inner products between pairwise data points in the high-dimensional quantum feature space (Havlíček et al. 2019; Schuld and Killoran 2019). The classification based on the quantum kernel method is known as quantum support vector machine (QSVM), which is a quantum analog of classical support vector machine (SVM) that has been used for a variety of machine learning tasks (Cortes and Vapnik 1995; Schölkopf and Smola 2002). An advantage of QSVM with certain feature maps for classically hard problems has been mathematically analyzed for the regime of fault-tolerant quantum computing (Liu et al. 2021; Jäger and Krems 2023). On the other hand, current quantum computers are still noisy intermediate-scale quantum (NISQ) devices (Preskill 2018); that is, NISQ processors are error-prone and error mitigation is sometimes necessary to reduce the impact of errors (Temme et al. 2017; LaRose et al. 2022). Despite the challenges, with the aid of cloud computing technology, there has been growing interest in the quest for early practical applications of near-term devices (Bharti et al. 2022).

In recent years, there has been remarkable progress in quantum hardware (de Leon et al. 2021), opening the path for the implementation of NISQ algorithms. Previous studies on quantum kernels have explored the use of various quantum hardware platforms, such as superconducting qubits (Havlíček et al. 2019; Djehiche and Löfdahl 2021; Heredge et al. 2021; Peters et al. 2021; Wang et al. 2021; Hubregtsen et al. 2022; Krunic et al. 2022), trapped-ion qubits (Moradi et al. 2022), Gaussian Boson Sampling (Schuld et al. 2020; Giordani et al. 2023), neutral-atom qubits (Albrecht et al. 2023), and nuclear-spin qubits (Kusumoto et al. 2021). Owing to quantum decoherence and the noise of quantum gates, one can typically perform a limited number of quantum operations on NISQ devices. In this regard, trapped-ion quantum processors seem to offer some advantages, thanks to long coherence time, all-to-all connectivity, and high-fidelity gate operations (Bruzewicz et al. 2019). Previous studies have demonstrated the implementation of different NISQ algorithms on trapped-ion quantum computers; in particular, researchers have recently used the IonQ Harmony quantum processor and reported interesting results in quantum machine learning (Johri et al. 2021; Ishiyama et al. 2022; Rudolph et al. 2022), finance (Zhu et al. 2022), quantum chemistry (Nam et al. 2020; Zhao et al. 2023), and the generation of pseudo-random quantum state (Cenedese et al. 2023). A recent study has shown the feasibility of implementing QSVM with a simple quantum circuit on a trapped-ion quantum computer (Moradi et al. 2022); nonetheless, further investigation is necessary to understand the full potential of the quantum kernel method on this platform using a different quantum kernel and various datasets.

In the present work, we investigate the performance of quantum support vector classification (QSVC) and quantum support vector regression (QSVR) on a trapped-ion quantum computer, using datasets from different industry domains including finance and materials science, aiming to bridge the gap between potential quantum computing applications and real-world industrial needs. Here, we employ quantum kernels described by a shallow quantum circuit that can be implemented on the IonQ Harmony quantum processor and analyze the performance of the models, in comparison with that of the classical counterpart as well as with that obtained from noiseless quantum-circuit simulations.

The remainder of the paper is organized as follows. To estimate the number of quantum measurements necessary for the estimation of quantum kernels for reliable predictions, we first perform noiseless and noisy quantum-computing simulations before conducting quantum experiments. Next, we investigate the effect of noise on the performance of the QSVC models using noisy simulations with various values for qubit-gate error rates. Then we report the results of QSVMs on the trapped-ion quantum processor. We train our QSVC models using a dataset containing fraudulent credit card transactions and image datasets such as the MNIST dataset and the Fashion-MNIST (Xiao et al. 2017) dataset. Also, we train our QSVR models using a financial market dataset and a dataset for superconducting materials. In the QSVR tasks, to reduce the effect of noise, we use a low-rank approximation of the noisy quantum kernel and carefully optimize hyperparameters in SVMs. We demonstrate that our quantum kernel can be used for both the QSVC and QSVR tasks for our datasets examined. Finally, we summarize our conclusions.

## 2 Results

### 2.1 Quantum circuit and the quantum kernel method

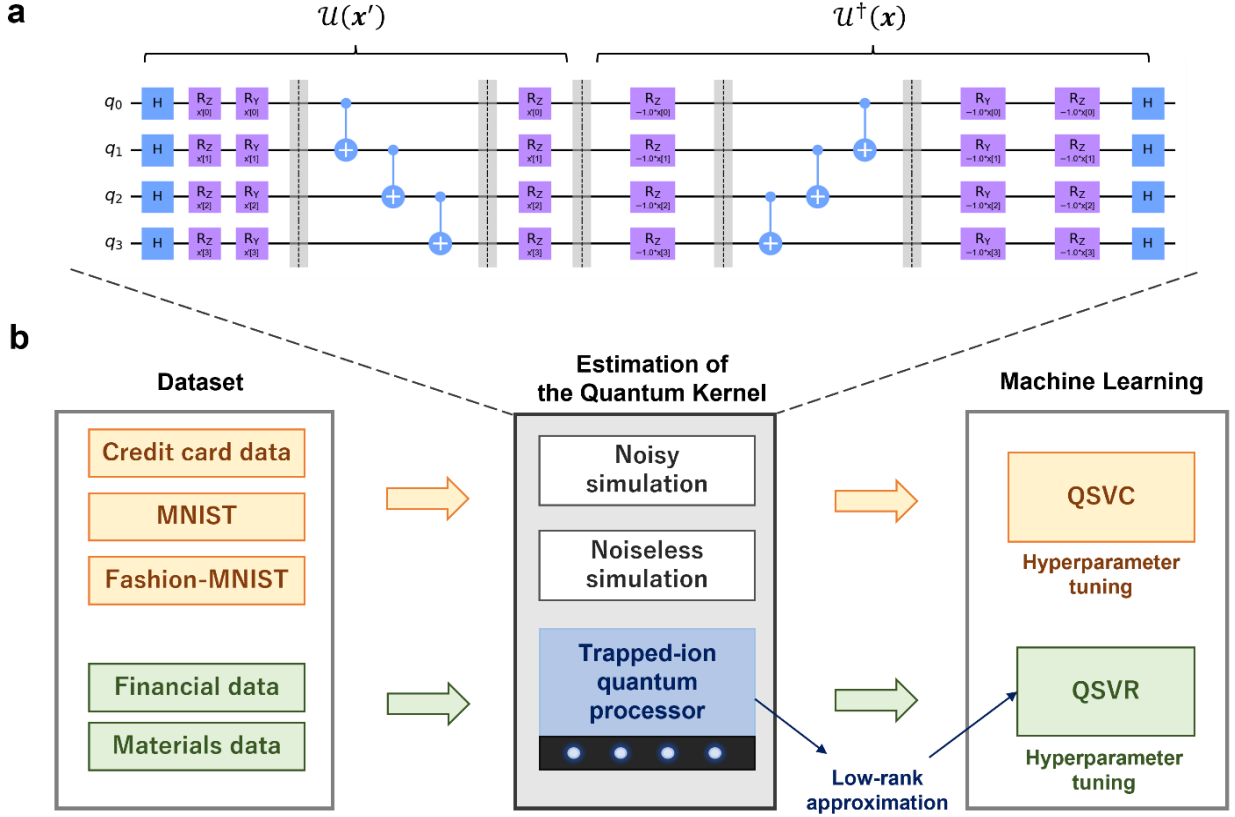

**Fig. 1** Schematic representation of our QSVC and QSVR workflows. (a) Shallow quantum circuit for estimating our quantum kernel $|\langle\phi(x)|\phi(x')\rangle|^2$. Our quantum feature map $|\phi(x)\rangle = \mathcal{U}(x)|0^{\otimes n}\rangle$ is given in Eq. 1 in the text. (b) Workflow for our QSVC and QSVR tasks (indicated by orange and green, respectively). For QSVC tasks, a credit-card dataset, the MNIST dataset, and the Fashion-MNIST dataset were used. For QSVR tasks, a financial dataset and a materials dataset were used. The estimation of the quantum kernel can be obtained using a quantum circuit simulator (either with or without noise) on a CPU (indicated by the white boxes) or computed using the IonQ Harmony (indicated by the blue box). Low-rank approximation was employed for QSVR tasks to reduce noise in the quantum kernel (for more details, see Sec. 2.3). The optimization of the hyperparameters in the models was also performed.

In the NISQ era, two-qubit gates are typically an order of magnitude lower in fidelity compared to single-qubit gates. This means that one can only perform a limited number of quantum operations to ensure that the results are distinguishable from noise. In the present study, we use the following quantum feature map using a shallow quantum circuit:

$$|\phi(x)\rangle = \mathcal{U}(x)|0^{\otimes n}\rangle = \left(\otimes_{q=1}^{n} R_z(x_q)\right) U_{2^n}^{\text{ent}} \left(\otimes_{q=1}^{n} \left(R_y(x_q) R_z(x_q) H\right)\right)|0^{\otimes n}\rangle,$$

and

$$U_{2^n}^{\text{ent}} := \prod_{q=1}^{n-1} \text{CNOT}_{q,q+1}.$$

(2)

Note that the connectivity of qubits in Eq. 1 is limited to their neighbors, resulting in $(n-1)$ interactions. This can make quantum computation more amenable for near-term quantum devices. The quantum feature map given in Eq. 1 has been applied to image classification using a specialized quantum-kernel simulator, which is highly customized for this particular quantum circuit using field programmable gate arrays (Suzuki et al. 2023). In quantum machine learning, the quantum kernel $K(x, x')$ described by the quantum feature map can be estimated by the inner product of the quantum states obtained from the two data points $x$ and $x'$:

$$K(x, x') = |\langle \phi(x)|\phi(x')\rangle|^2 = \left|\langle 0^{\otimes n}|U^\dagger(x)U(x')|0^{\otimes n}\rangle\right|^2.$$

(3)

The kernel represents the similarity between the two data points in the high-dimensional Hilbert space. The quantum kernel entry can be estimated by using a shallow quantum circuit described in Fig. 1a. Once the quantum kernel is estimated by a quantum computer or a quantum-circuit simulator, we can use the kernel-based method (Fig. 1b). The goal of SVM is to find the decision function for binary classification. Suppose we are given a set of samples $(x_i, y_i)$ with $x_i \in \mathbb{R}^d$ and $y_i \in \{\pm 1\}$, where $d$ is the dimension of the input vector and the index $i$ runs over $1, \cdots, m$. To find the decision function, we solve the following problem (Schölkopf and Smola 2002):

$$\max_{\alpha \in \mathbb{R}^m} W(\alpha) = -\frac{1}{2} \sum_{i,j=1}^{m} \alpha_i \alpha_j y_i y_j K(x_i, x_j) + \sum_{i=1}^{m} \alpha_i,$$

(4)

subject to

$$\alpha_i \in [0, C] \text{ and } \sum_{i=1}^{m} y_i \alpha_i = 0.$$

(5)

Here, the coefficients $\{\alpha_i\}$ are parameters determined through the optimization process. The patterns $x_i$ for which $\alpha_i > 0$ are called *support vectors* (SVs). The regularization parameter $C$ controls the tradeoff between model complexity and its capacity to tolerate errors. The decision function $f(x)$ takes the form:

$$f(x) = \text{sgn}\left(\sum_{i=1}^{m} y_i \alpha_i K(x_i, x_j) + b\right).$$

(6)

The bias $b$ can be determined by the support vectors once the SVs and their Lagrange multipliers are obtained by the dual optimization (Schölkopf and Smola 2002).

## 2.2 Noiseless and noisy quantum simulations

To understand the effects of noise, we first employed a device-noise model provided by Qiskit Aer (Aleksandrowicz et al. 2019). Here, the noise model is based on a depolarizing noise model, in which single-qubit gate errors and two-qubit gate errors are taken into account. Single-qubit errors consist of a single-qubit depolarizing error followed by a single-qubit thermal relaxation error, whereas two-qubit gate errors comprise a two-qubit depolarizing error followed by single-qubit thermal relaxation errors on both qubits in the gate. Hereafter, we denote them as $p_1$ and $p_2$, respectively. In the context of the quantum kernel method, it is of particular importance to understand how noise affects the quality of the quantum kernel matrix and the prediction accuracy. To quantify this, we use the *alignment* between two kernels (Cristianini et al. 2001) defined by

$$A(K, K') = \frac{\langle K, K' \rangle_F}{\sqrt{\langle K, K \rangle_F \langle K', K' \rangle_F}},$$

(7)

where $\langle P, Q \rangle_F$ is the Frobenius inner product between the matrices $P$ and $Q$:

$$\langle P, Q \rangle_F = \sum_{ij} P_{ij} Q_{ij} = \mathrm{Tr}\{P^T Q\}.$$

(8)

The alignment $A(K, K')$ can be viewed as the cosine of the angle between the two matrices viewed as vectors. By using the alignment of the noisy quantum kernel $K^{\mathrm{noise}}$ with the noiseless quantum kernel $K$, $A(K, K^{\mathrm{noise}})$, we can conveniently measure the deviation of a noisy kernel from the noiseless one.

Using device-noise-model simulations, we investigated the robustness of our quantum kernel matrix in the presence of noise and the prediction performance (Fig. 2). In our noisy simulations, we varied the number of qubits from 4 to 12 in Eq. 1 and considered the following conditions for qubit-gate error rates: (i) $p_1 = 0.001$, $p_2 = 0.005$ and (ii) $p_1 = 0.01$, $p_2 = 0.05$. In Appendix A, we numerically show that 500 shots per kernel entry were enough to ensure the quality of our quantum kernel and to maintain reliable predictions; thus, 500 shots were conducted for each kernel entry throughout our simulations. To explore the applicability of our quantum kernel, three different datasets were considered: the credit card fraudulent transaction dataset, the NMIST dataset, and the Fashion-MNIST dataset. For the three datasets, the test accuracy obtained from the noisy quantum kernel was on par with that obtained from the noiseless quantum simulations, suggesting that the noise in the quantum kernel had minimal impact on the test accuracy of our QSVM models. This can be confirmed by the fact that the alignment was above 0.996, which may be partly due to the nature of our shallow quantum circuits. On the other hand, we are aware that our simulations based on the device noise model are only an approximation of real errors that occur on actual devices (In Section 2.3, we will demonstrate the performance of our QSVC models on the real quantum device using 4 qubits).

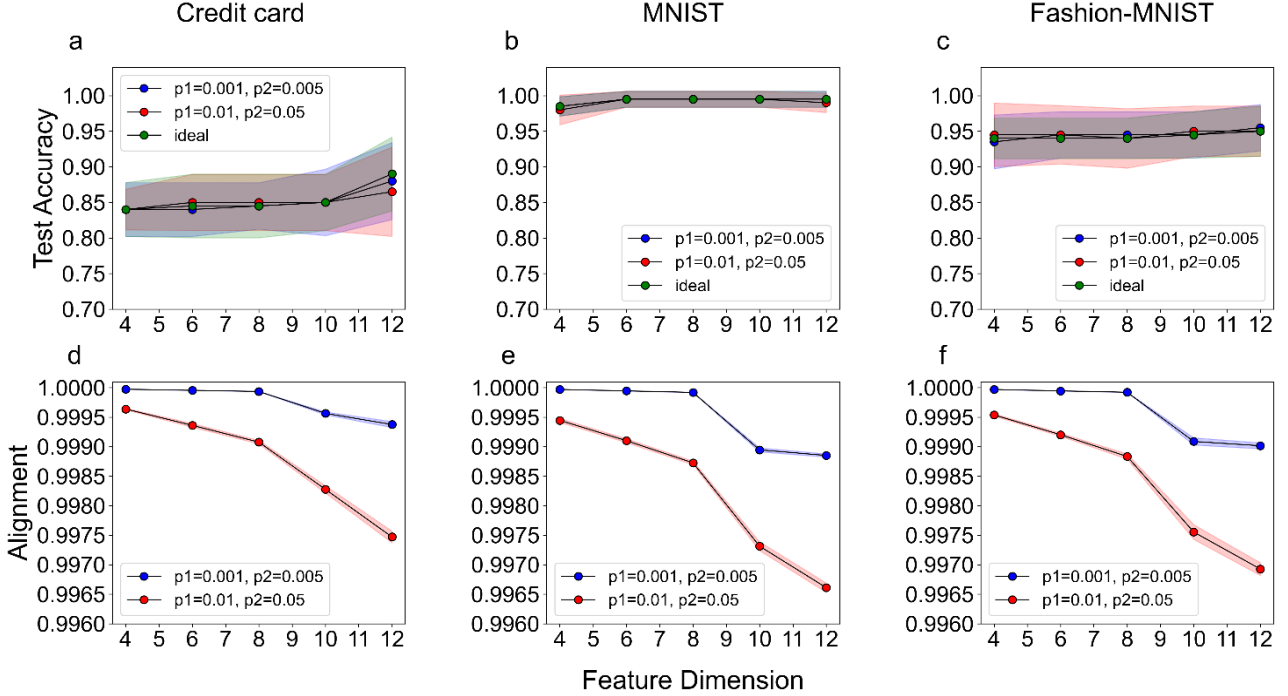

**Fig. 2** The dependence of the prediction performance of our QSVC models on the number of qubits from 4 to 12 (noisy simulations). The error bars indicate the standard deviation obtained from 5 independent seeds. Top panel: test accuracy for (a) credit card dataset, (b) MINST dataset (binary classification of two labels: '0' vs. '1'), and (c) Fashion-MNIST dataset (binary classification of two image categories: 'T-shirt' vs. 'trouser'). Bottom panel: alignment of the noisy quantum kernel with the noiseless quantum kernel for (d) credit card dataset, (e) MINST dataset, and (f) Fashion-MNIST dataset. In our device-noise-model simulations, we consider the following conditions for single- and two-qubit gate error rates: (i) $p_1 = 0.001$, $p_2 = 0.005$ (indicated by blue); (ii) $p_1 = 0.01$, $p_2 = 0.05$ (indicated by red). Five independent seeds for each dataset were used to obtain the statistical results. The number of training data was 40 and the number of test data was 20.

Next, we investigated how the device noise level affects the alignment and the test accuracy (Fig. 3). To this end, we performed device-noise-model simulations using 4 qubits for a range of qubit-gate error rates: $0.001 \leq p_1 \leq 0.55$ and $0.001 \leq p_2 \leq 0.55$. By comparing the alignment and the test accuracy, we found that, for certain regions ($p_1 < \approx 0.05$ and $p_2 < \approx 0.1$), our QSVC model can predict, even in the presence of noise (Fig. 3a and 3b). When qubit-gate error rates become relatively high, however, the noisy quantum kernel deviates from the ideal quantum kernel and thus the prediction performance begins to deteriorate rapidly. Given the fact that $p_1 \approx 0.001$ and $p_2 \approx 0.01$ for state-of-the-art NISQ devices, we can validate the prediction performance of our QSVC model on real quantum computers (which will be demonstrated in the next subsection). Our noise-model simulations suggest that the alignment between noiseless and noisy quantum kernels is an indicator of how reliably a QSVC model can predict using a NISQ device in comparison with its noiseless counterpart. In our QSVC model, if the alignment is higher than 0.98, then the QSVC model can make reliable predictions (Fig. 3c). To understand this more intuitively, one can recall that the alignment can be viewed as the cosine of the angle between the two matrices (viewed as vectors); in such a mathematical viewpoint, it means that the angle between noiseless and noisy kernels needs to be less than 11.5 degrees for reliable predictions on a noisy quantum device.

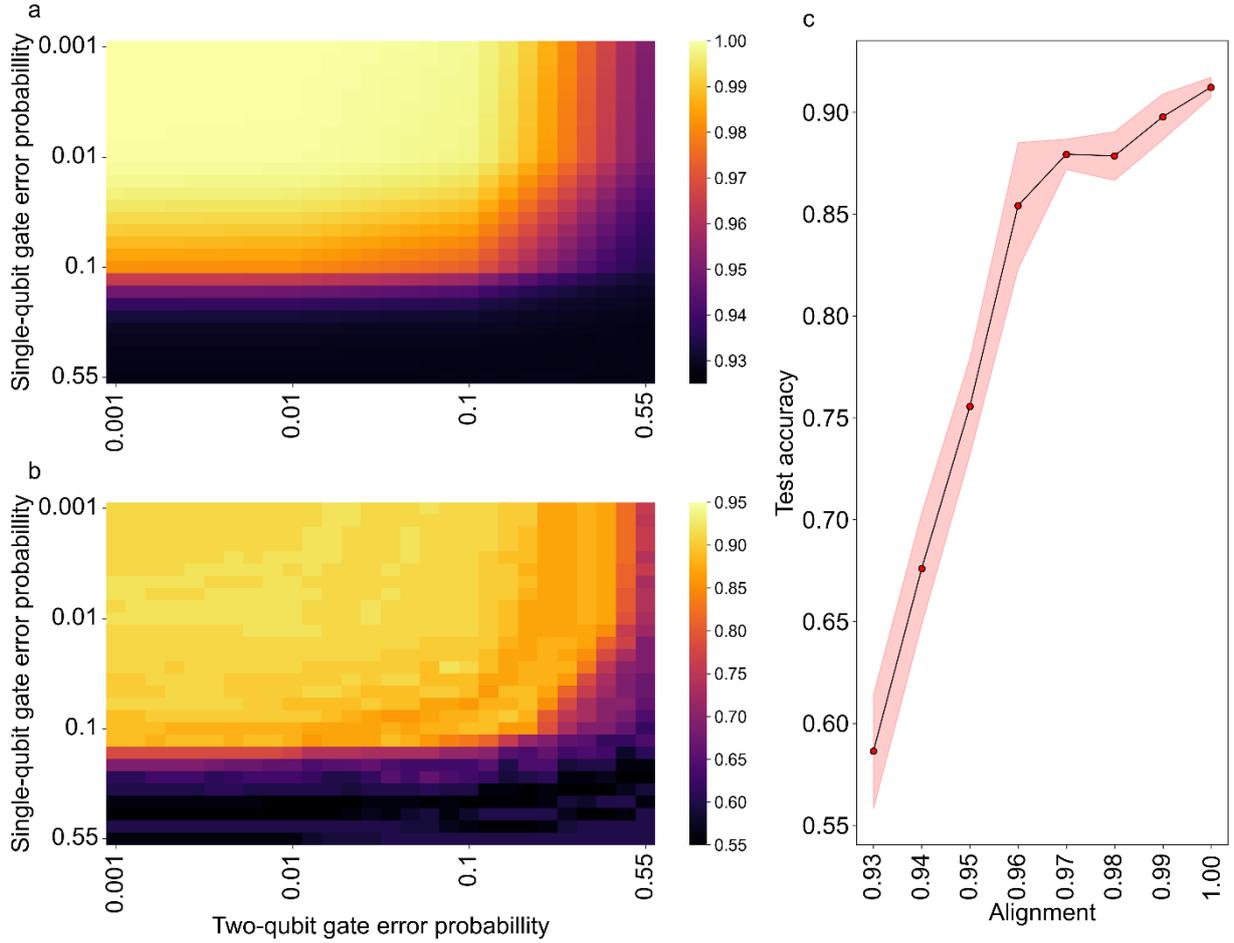

**Fig. 3** Effects of noise on our QSVM model. (a) the alignment, (b) the test accuracy of the QSVC model, and (c) the correlation between the two. The device-noise-model simulations were performed using 4 qubits for a range of qubit-gate error rates: $0.005 \leq p_1 \leq 0.55$ and $0.005 \leq p_2 \leq 0.55$. The results suggest that our QSVC model is capable of making reliable predictions, if the alignment is higher than 0.98, which roughly corresponds to the condition that $p_1 < \approx 0.05$ and $p_2 < \approx 0.1$. The shaded area in part (c) indicates the standard deviation. The Fashion-MNIST dataset was used, and the number of training data was 40 and the number of test data was 20.

## 2.3 QSVC on the IonQ Harmony quantum computer

Having examined the results of the noise-model simulations, we now turn to our quantum experiments using the IonQ Harmony. The Gram matrices we obtained using the quantum device (4 qubits) are shown in Fig. 4. To validate the quality of our noisy quantum kernels, we investigated the alignment of the noisy quantum kernel with the noiseless quantum kernel: the values for the alignment $A(K, K^{\text{noise}})$ were 0.986, 0.984, and 0.993 for the credit-card dataset, the MNIST dataset, and the Fashion-MNIST dataset, respectively. Since the three values were higher than 0.98, this suggests that the quantum kernel matrix entries were successfully estimated using the IonQ Harmony and indicates that reliable predictions can be made using our QSVC models on the quantum device (see also Fig. 3c).

Motivated by the reliable estimation of the quantum kernel on the IonQ Harmony, we trained QSVC models using the three datasets and validated the models using test data (Table 1). For comparison, we used classical Gaussian kernels $K(x, x') = \exp(-\gamma \|x - x'\|^2)$. For the credit card dataset, the classical SVM parameters were the regularization constant $C = 3.2$ and $\gamma = 0.25$, whereas QSVM parameters were $C = 6.2$ for the noiseless simulation and $C = 4.2$ for the IonQ machine. For the MNIST dataset, the classical SVM had $C = 3.5$ and $\gamma = 0.25$, whereas both the noiseless and noisy QSVMs had $C = 1.0$. Finally, for the Fashion-MNIST dataset, the classical SVM parameters were $C = 1.5$ and $\gamma = 0.25$, whereas the QSVM had $C = 0.4$ for the noiseless simulation and $C = 1.0$ for the IonQ machine.

Our results show that the prediction performance of our QSVC models was maintained even in the presence of noise. Test accuracies achieved with the quantum computer for the credit-card dataset, the MNIST dataset, and the Fashion-MNIST dataset were 70%, 100%, and 100%, respectively, reflecting equivalent performance to the QSVC models using noiseless quantum kernels. This is consistent with the results of our noise-model simulations, in which predictions can be made using noisy quantum kernels with an alignment higher than 0.98. Furthermore, the performance of our QSVC models on the IonQ Harmony was comparable to that of the classical counterparts. In Supplementary Information, we also included the results of the IonQ Aria experiments involving 4 and 8 qubits. Thanks to the improved fidelity of the two-qubit gate operations in the IonQ Aria, we were able to obtain a quantum kernel using an 8-qubit system. Our QSVC model with 8 qubits achieved a 100% test accuracy on the Fashion-MNIST dataset. On a final note, we mention the number of support vectors (note that the decision boundary for the largest margin is determined solely by the position of the support vector): we found that there was a slight difference in the number of support vectors between QSVC models with noiseless kernels and those with noisy kernels on the IonQ Harmony, which might imply a subtle difference in the quantum feature map between noiseless simulations and actual quantum experiments, though both the QSVC models gave the same test accuracies.

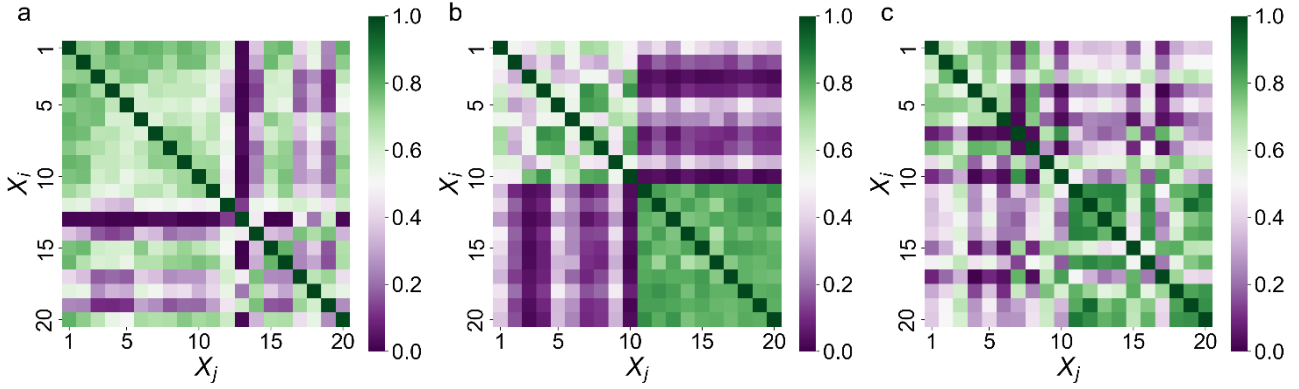

**Fig. 4** Quantum kernel matrices obtained using the IonQ Harmony quantum computer. (a) Credit-card dataset, (b) MNIST dataset, and (c) Fashion-MNIST dataset. For all the cases, 4 qubits were used for obtaining the matrices, with the number of training data $N = 20$. The values for the alignment of the noisy quantum kernel with the noiseless quantum kernel, $A(K, K')$, were 0.986, 0.984, and 0.993 for (a), (b), and (c), respectively. All of the values were higher than 0.98, which suggests that the quantum kernel matrix entries were successfully obtained using the IonQ quantum computer and indicates that reliable predictions can be made using our QSVC models on the quantum device (see also Fig. 3c).

**Table 1** Model accuracy and the number of support vectors (SVs) for classical and quantum SVMs on three different datasets. The dimension for the input data was reduced to 4.

| Dataset | Model | Number of SVs | Acc. (train) (%) | Acc. (test) (%) |
|---|---|---|---|---|
| Credit card | | | | |
| | Classical SVM | 12 | 80 | 70 |
| | QSVM (noiseless sim.) | 11 | 85 | 70 |
| | QSVM (IonQ Harmony) | 12 | 100 | 70 |
| MNIST | | | | |
| | Classical SVM | 16 | 100 | 100 |
| | QSVM (noiseless sim.) | 6 | 100 | 100 |
| | QSVM (IonQ Harmony) | 10 | 100 | 100 |
| Fashion-MNIST | | | | |
| | Classical SVM | 8 | 100 | 100 |
| | QSVM (noiseless sim.) | 18 | 100 | 100 |
| | QSVM (IonQ Harmony) | 15 | 100 | 100 |

### 2.4 QSVR on the IonQ Harmony quantum computer

*Datasets.* Two different datasets were used in our QSVR tasks. One is a financial-market dataset (given in Tables S2 and S3 in Supplementary Information). Financial data are characterized by high volatility and are often subject to noise caused by random fluctuations. In recent years, pandemics, geopolitical risks, and other microeconomic factors have caused supply chain disruption, which led to price fluctuations of metal commodities such as nickel. In our QSVR model, the target variable $y_i$ was the UK nickel price, and three attributes $x_i \in \mathbb{R}^3$ were considered: the Shanghai Stock Exchange Composite (SSE) Index, West Texas Intermediate (WTI) crude oil, and US dollar index. Thus, in our QSVR model, 3 qubits were used for describing the quantum feature map on the Ion Q Harmony. Hereafter, the dataset is referred to as the *financial* dataset. The other is a superconducting-materials dataset (Hamidieh 2018). Here, the target variable $y_i$ was the critical temperature $T_c$ for a broad class of superconducting materials. The original dataset contains 81 features (or descriptors); by using dimensionality reduction, four-dimensional vectors $x_i \in \mathbb{R}^4$ were used as input data in this work (for more details of preprocessing, see section 4.1). Hence, 4 qubits were used to describe the quantum feature map. Hereafter, the dataset is referred to as the *materials* dataset.

*Low-rank approximation in the noisy quantum kernel.* A popular approach to reducing the noise in the quantum kernel is to use a depolarizing model (Hubregtsen et al. 2022; Moradi et al. 2022); however, such a noise model may not necessarily be suited for real quantum devices, because there are various sources of noise. In addition, at the time of conducting our quantum experiment, we were not able to access the full control of native quantum gates of the trapped-ion quantum computer in the cloud service. In this work, we rather employed a postprocessing approach for error mitigation; in particular, we used low-rank approximation to reduce the noise in the quantum kernel. This can maintain the important information of the original matrix while reducing the noise. The low-rank approximation can be performed, for instance, by singular value decomposition (SVD). Recently, a study by Wang et al. (2021) showed that the training performance of noisy quantum kernels is improved when spectral transformation (eigendecomposition) is adopted. Our idea is to reconstruct a quantum kernel $\widehat{K}$ from a noisy quantum kernel $K$ by using eigendecomposition:

$$\widehat{K} = \sum_{k=1}^{r} \mu_k \boldsymbol{u}_k \boldsymbol{u}_k^{\mathrm{T}},$$

(9)

where $\mu_k$ is the $k$th eigenvalue and $\boldsymbol{u}_k$ is the corresponding $k$th eigenvector. The quantum kernel is approximated by summing over $\mu_k \boldsymbol{u}_k \boldsymbol{u}_k^{\mathrm{T}}$, where the index $k$ runs over $1, \cdots, r$. Motivated by the important role of the alignment in QSVC (Fig. 2c), we argue that the optimal value for $r$ can be determined by maximizing the alignment of the noisy quantum kernel with the noiseless one:

$$r^* = \underset{r \in \mathbb{N}}{\operatorname{argmax}} A(K, \widehat{K}).$$

(10)

In the case of test data, we calculate a train-test kernel matrix, which is generally a rectangular matrix; hence, SVD was used for low-rank approximation.

We investigated the effects of low-rank approximation in improving the quality of the noisy quantum kernel (Fig. 5). For the financial dataset, the alignment had the maximum value of 0.993 at $r^* = 8$ (Fig. 5a). For the materials dataset, on the other hand, the alignment had the maximum value of 0.984 at $r^* = 10$ (Fig. 5d). The difference in the optimal $r^*$ appears to be related to the difference in the nature of the datasets. For both cases, after the alignment peaked at the optimal $r^*$, the value for the alignment was gradually decreased and finally saturated for larger values of $r$. This can be confirmed by the fact that the contribution of eigenvectors for $k > r^*$ became substantially small (Fig. 5b and 5e), indicating that a large portion of the information is concentrated in eigenvectors up to $r^*$th. The results suggest that low-rank approximation can improve the quality of the noisy quantum kernel to some extent.

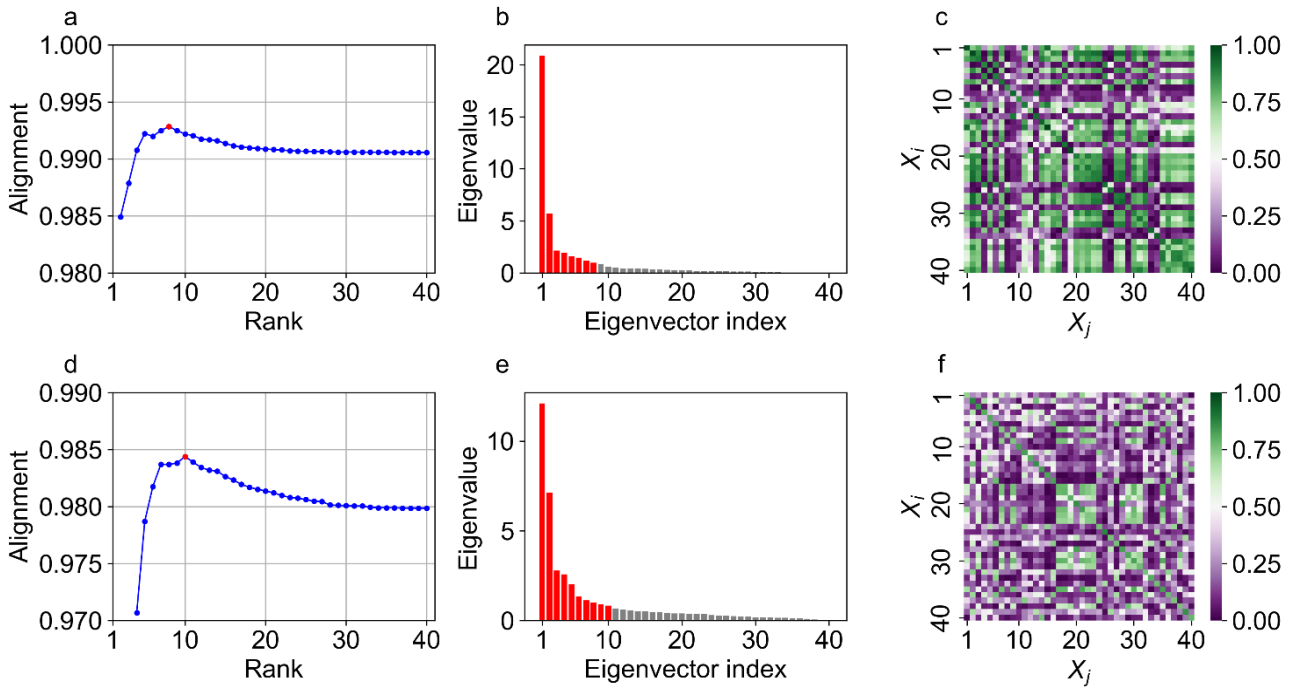

**Fig. 5** Use of spectral decomposition in improving the quantum kernel on the quantum processor. Top (financial dataset; 3 qubits): (a) alignment between the reconstructed quantum kernel and the noiseless quantum kernel with respect to the rank in the low-rank approximation (the alignment has the maximum value of 0.993 at $r^* = 8$, which is indicated by the red circle); (b) eigenvalues with respect to the eigenvector index (the first 8 components are indicated by the red bars); (c) reconstructed quantum kernel using the low-rank approximation ($r^* = 8$). Bottom (materials data; 4 qubits): (d) alignment between the reconstructed quantum kernel and the noiseless quantum kernel with respect to the rank in the low-rank approximation (the alignment has the maximum value of 0.984 at $r^* = 10$, which is indicated by the red circle); (e) eigenvalues with respect to the eigenvector index (the first 10 components are indicated by the red bars); (f) reconstructed quantum kernel using the low-rank approximation ($r^* = 10$).

*Optimization of hyperparameters in $\varepsilon$-support vector regression (SVR).* The goal of $\varepsilon$-SVR is to find a regression function $f(x)$ that has at most $\varepsilon$ deviation from the obtained targets $\{y_i\}$ for all the training data $\{x_i\}$ (Schölkopf and Smola 2002). Here, the hyperparameter $\varepsilon$ defines a margin of tolerance (or $\varepsilon$- insensitive tube) where no penalty is associated with errors. In other words, any data points within this allowable error range are not considered errors, even if they do not fall directly on the regression line. This can be realized by using the $\varepsilon$-insensitive loss function introduced by Vapnik, which is an analog of the soft margin in SVC (Schölkopf and Smola 2002). The linear $\varepsilon$-insensitive loss function can be described by

$$L_\varepsilon = \begin{cases} 0 & |y - f(x)| \leq \varepsilon \\ |y - f(x)| - \varepsilon & \text{otherwise} \end{cases}.$$

(11)

A smaller value of $\varepsilon$ narrows the 'no penalty' region, making the model more sensitive to the training data, whereas a larger value of $\varepsilon$ creates a wider tube, making the model less sensitive to the training data. An appropriate value of $\varepsilon$ is related to the noise magnitude of data (Zhang and Han 2013). By using the Lagrangian formalism and introducing a dual set of variables, the primal problem in $\varepsilon$-SVR can be transformed into the dual optimization problem (Schölkopf and Smola 2002), which is described as follows:

$$\max_{\boldsymbol{\alpha}, \boldsymbol{\alpha}^* \in \mathbb{R}^m} W(\boldsymbol{\alpha}, \boldsymbol{\alpha}^*) = -\frac{1}{2} \sum_{i,j=1}^{m} (\alpha_i^* - \alpha_i)(\alpha_j^* - \alpha_j) K(\boldsymbol{x}_i, \boldsymbol{x}_j) - \varepsilon \sum_{i=1}^{m} (\alpha_i^* + \alpha_i) + \sum_{i=1}^{m} (\alpha_i^* - \alpha_i) y_i,$$

(12)

subject to

$$\alpha_i^*, \alpha_i \in [0, C] \text{ and } \sum_{i=1}^{m}(\alpha_i^* - \alpha_i) = 0.$$

(13)

Here, the coefficients $\{\alpha_i\}$ and $\{\alpha_i^*\}$ are parameters determined through the optimization process. The regularization parameter $C$ determines the tradeoff between the complexity of the model and its capacity to tolerate errors. A larger value of $C$ makes the model less tolerant of errors, which potentially leads to a risk of overfitting, whereas a smaller value of $C$ helps the model more tolerant of errors, which tends to make the model less complex. By tuning the hyperparameters $\varepsilon$ and $C$, one can find a good combination of parameters that makes the model more robust on new data, thus improving its generalization performance. The regression function takes the form (Schölkopf and Smola 2002):

$$f(\boldsymbol{x}) = \sum_{i=1}^{m} (\alpha_i^* - \alpha_i) K(\boldsymbol{x}_i, \boldsymbol{x}) + b.$$

(14)

The bias $b$ can be determined by the SVs, once the Lagrange multipliers are obtained by the dual optimization. To access the performance of the model, we used the root mean square error (RMSE):

$$\text{RMSE} = \sqrt{\frac{1}{N}\sum_{i=1}^{N}(y_i - \hat{y}_i)^2}.$$

(15)

By performing a grid search for $\varepsilon$ and $C$, we optimized the hyperparameters in $\varepsilon$-SVR using the quantum kernel that had been reconstructed by low-rank approximation (Fig. 6). For the financial dataset, the optimal values for the hyperparameters $\varepsilon$ and $C$ were 0.21 and 1.4, respectively, indicating that the $\varepsilon$-insensitive loss function was effective in enhancing the generalization of the model and reducing overfitting. The result is partly due to the nature of the financial market data; that is, the data is characterized by high volatility owing to various factors such as geopolitical events and market sentiment and is often subject to noise caused by random fluctuations. For the materials dataset, on the other hand, the optimal values for the hyperparameters $\varepsilon$ and $C$ were 0.0 and 0.3, respectively. The optimal value of $\varepsilon = 0$ means that the allowable error range in the training data was unnecessary for this particular case. A possible reason may be that the impact of noise in the quantum kernel was effectively canceled through dimensionality reduction of input data and the use of low-rank approximation of the noisy quantum kernel. At the same time, a slightly smaller value of $C = 0.3$ made the model more robust against overfitting. For both cases, increasing the value for $C$ (i.e., making the model fit the training data more tightly) did not improve the performance; instead, it had an adverse effect on the results (Fig. 6). Our results suggest that a combined approach that involves both the low-rank approximation to the noisy quantum kernel and the optimization of the hyperparameters in $\varepsilon$-SVR can be a useful strategy for improving the performance and robustness of the QSVR models.

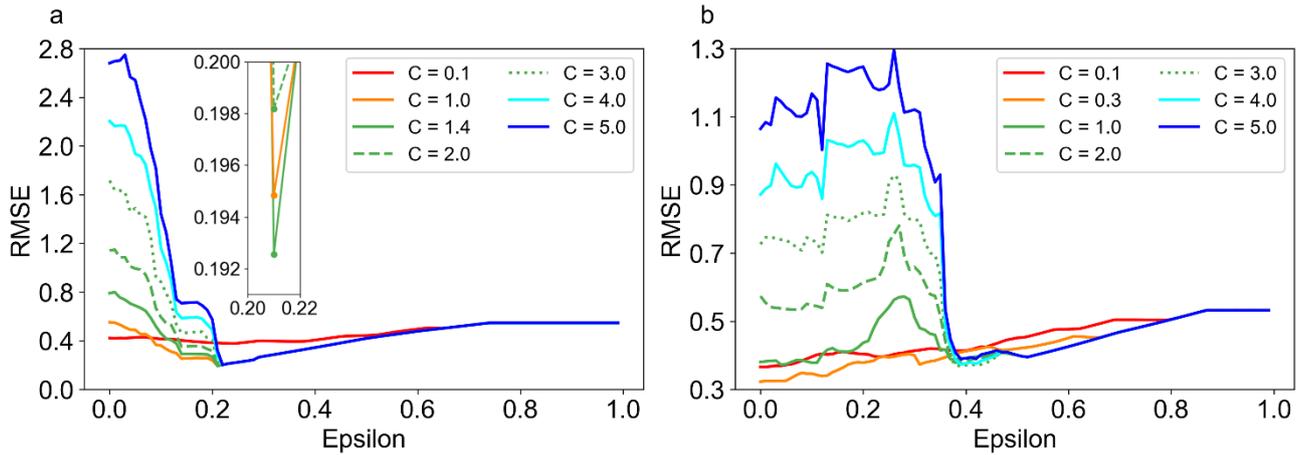

**Fig. 6** Optimization of the hyperparameters $C$ and $\varepsilon$ in $\varepsilon$-SVR using the quantum kernel reconstructed by low-rank approximation (see also Fig. 5). RMSE with respect to $\varepsilon$ for (a) financial dataset and (b) materials dataset. The optimal values for $(\varepsilon, C)$ were (0.21, 1.4) and (0.0, 0.3) for (a) and (b), respectively.

*Performance.* Herein, we report the results of our QSVR models on the IonQ trapped-ion quantum computer and compare the performance with that obtained by the classical SVR tasks. For the classical SVR, we used Gaussian kernels $K(x, x') = \exp(-\gamma \|x - x'\|^2)$. The optimized hyperparameters ($\gamma, \varepsilon, C$) were (0.6, 0.1, 5.7) and (0.3, 0.18, 3.4) for the financial dataset and the materials dataset, respectively. For the financial dataset, the performance of our QSVR model using the noiseless simulation was comparable to that of our classical SVR model: the coefficient of determination ($R^2$) for the classical SVR model was 0.930; whereas that for the QSVR model was 0.932. On the other hand, $R^2$ for the QSVR model using the IonQ Harmony was 0.868, which was 6.9% lower than that obtained by the noiseless simulation. The QSVR model worked well in predicting the financial price for this particular period (note that the model may not guarantee a similar performance for another time window). For the materials dataset, $R^2$ for the classical SVR model was 0.728, whereas that for the QSVR with the noiseless quantum kernel was 0.703. The coefficient of determination $R^2$ for the QSVR model using the real quantum device was 0.628, which was 10.7% lower than that obtained from the noiseless simulation. Since the alignment of the reconstructed quantum kernel for the materials dataset (0.984) was slightly lower than that for the financial dataset (0.993), the decrease in $R^2$ observed in the real device appeared to be more significant in the case of the former dataset. Overall, the presence of noise negatively affected the performance of QSVR tasks. In QSVC models, the performance appears to be less impacted by noise, as data points can be effectively separated in the high-dimensional space. However, regression models, which predict real values, were more acutely affected by noise. Similar to the case with the QSVC models, the results suggest that the alignment of the quantum kernel is a reliable measure for accessing the performance of the quantum kernel method on the real quantum device.

**Table 2** Coefficients of determination $R^2$ and RMSE for classical and quantum SVR models on the financial dataset and the superconducting-materials dataset (for more details, see the text)

| Dataset | Model | $R^2$ | RMSE |
|---|---|---|---|
| Financial data | | | |
| | Classical SVR | 0.930 | 0.140 |
| | QSVR (noiseless sim.) | 0.932 | 0.138 |
| | QSVR (IonQ Harmony) | 0.868 | 0.192 |
| Materials data | | | |
| | Classical SVR | 0.728 | 0.276 |
| | QSVR (noiseless sim.) | 0.703 | 0.288 |
| | QSVR (IonQ Harmony) | 0.628 | 0.322 |

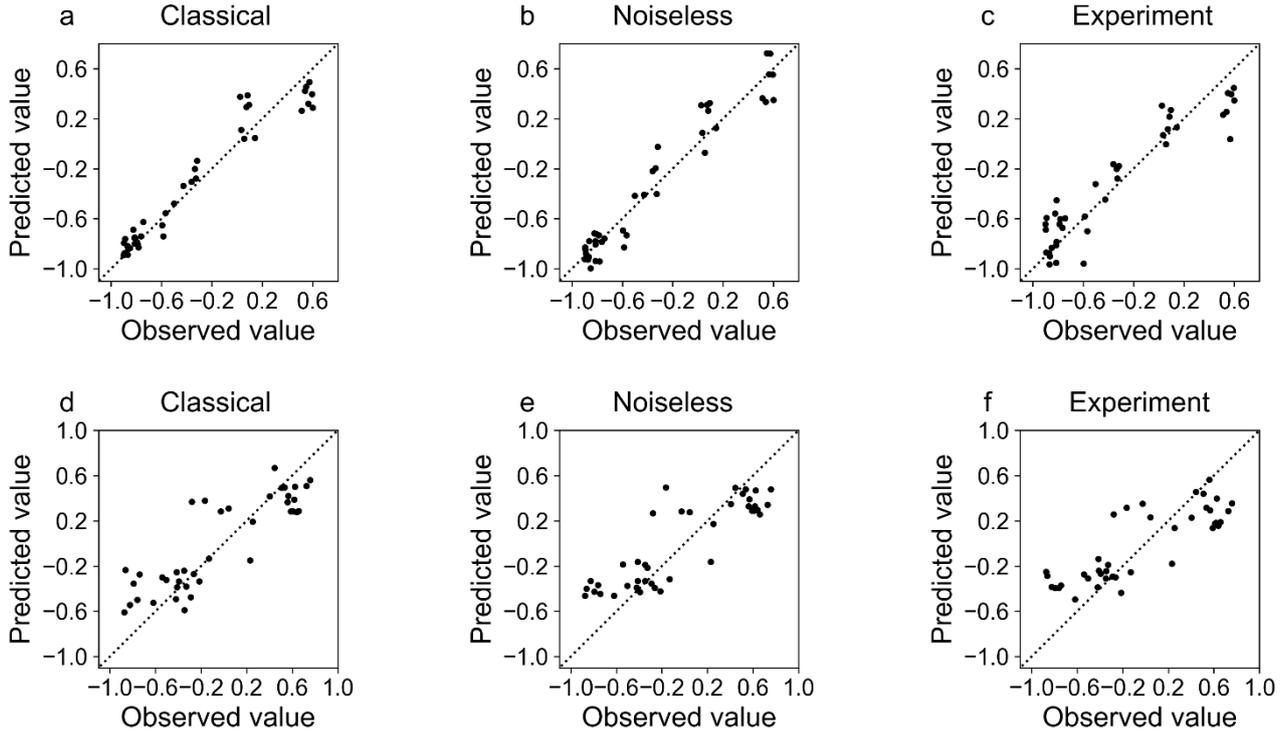

Fig. 7 Parity plot between the predicted and observed values. Top panel (financial dataset): (a) classical SVR ($R^2 = 0.930$); (b) QSVR using the noiseless simulation ($R^2 = 0.932$); (c) QSVR using the IonQ Harmony with 3 qubits ($R^2 = 0.868$). Bottom panel (materials dataset): (d) classical SVR ($R^2 = 0.728$); (e) QSVR using the noiseless simulation ($R^2 = 0.703$); (f) QSVR using the IonQ Harmony with 4 qubits ($R^2 = 0.628$). For all cases, the number of test data was 40.

## 3 Discussion

In the present work, we have investigated our QSVC and QSVR models, by performing quantum-circuit simulations and using the IonQ Harmony quantum processor. For the classification tasks, we used the credit card dataset, the MNIST dataset, and the Fashion-MNIST dataset. The performance of our QSVC models obtained using 4 qubits of the trapped-ion quantum computer was comparable to that of the classical counterparts and that of the QSVC models obtained from the noiseless quantum kernels. This suggests that the presence of noise in the quantum kernel had a minimal impact on the test accuracy of the QSVM models. Our quantum experiments with 4 qubits were consistent with the analysis of our device-noise simulations, in which the prediction performance can be maintained so long as the device noise level is lower than a certain threshold. The robustness of our quantum kernel in the presence of noise can be explained by the fact that the alignment between the noiseless and the noisy quantum kernels was close to one (the alignment was higher than 0.98). Hence, our results suggest that the alignment is a reliable measure for evaluating the performance of a QSVM model on a NISQ device in comparison with a noiseless counterpart.

In the case of our QSVR models, we used the financial dataset and the dataset for superconducting materials. In particular, we investigated the role of the low-rank approximation and the effects of the hyperparameter tuning in $\varepsilon$-SVR in improving the performance and robustness of the QSVR models. We found that the low-rank approximation was effective in reducing the effects of noise in the quantum kernel. The optimization of the hyperparameters in $\varepsilon$-SVR was also beneficial for mitigating the effect of noise. Therefore, a combined approach using the low-rank approximation to the noisy quantum kernel and the hyperparameter tuning in $\varepsilon$-SVR can be a useful method for enhancing the performance of the QSVR models. We have demonstrated that the quantum kernel described by our shallow circuit was versatile for both the QSVM and QSVR tasks for the different datasets we examined. While our quantum feature map did not necessarily exemplify a so-called quantum advantage because of its shallow quantum circuit and the limited number of qubits, our findings could provide valuable insights for designing quantum feature maps.

Let us now discuss open questions and challenges. A recent theoretical study by Thanasilp et al. (2022) shows that under certain conditions, quantum kernel entries can exponentially concentrate around a certain value (with an increasing number of qubits). On the other hand, such exponential concentration was not observed in the quantum experiments conducted for our tasks. This discrepancy can be attributed partly to our choice of shallow quantum circuits, which circumvent high expressibility, and partly to the nature of the datasets we used. In general, quantum kernels with high expressibility can lead to training difficulties. For instance, quantum kernels with $L$ layers of hardware-efficient circuit unit reach the exponential decay regime when $L$ is sufficiently large ($L \geq 75$); in contrast, for a small number of layers ($L \leq 8$), quantum kernels do not enter the exponential concentration regime (see Fig. 4 in the paper by Thanasilp et al. (2022)). Our quantum kernels consisting of low-depth circuits correspond to the latter scenario, which is in line with our successful results in training and prediction. Furthermore, noise can significantly impose limitations on the potential of quantum computing in the NISQ era. The presence of noise results in the loss of information during gate operations before extracting information through measurements. Although our quantum experiments have shown that training and prediction were feasible for systems with 4 and 8 qubits, an increase in the number of qubits (which in turn leads to an increase in two-qubit gate operations) may result in a substantial decrease in the values of quantum kernels, thereby making the training process more difficult. Lastly, the scalability of quantum kernels remains an open question (Thanasilp et al. 2022; Jerbi et al. 2023) and further advancements are necessary for the practical application of quantum kernel methods. In this context, a new field of geometric quantum machine learning (Meyer et al. 2023; Ragone et al. 2023; West et al. 2023), given its broad theoretical scope, could facilitate the development of carefully designed quantum kernels.

## 4 Computational details

### 4.1 Quantum-computing experiments

All the quantum calculations were carried out using the IonQ Harmony quantum processor provided by the Amazon Braket cloud service. To conduct our quantum-computing experiments, we have developed our quantum software development kit (SDK) called `PhiQonnect`, which is especially intended for quantum-kernel-based methods, including QSVC and QSVR. The quantum SDK utilizes open-source libraries such as IBM `Qiskit` (Aleksandrowicz et al. 2019) (ver. 0.39.5), `pytket` (ver. 1.11.1), which is a language-agnostic optimizing compiler provided by Quantinuum (Sivarajah et al. 2020), `amazon-braket-sdk` (ver. 1.35.3) developed by Amazon Braket (Amazon Web Services 2022), and `scikit-learn` (Pedregosa et al. 2011) (ver. 1.2.1), in which `LIBSVM` library (Chang and Lin 2011) is included. All the quantum computations on the NISQ device were obtained using our SDK, which is available as open-source software (see Code Availability).

The computational details of our quantum experiments using the IonQ Harmony are summarized in Table 3. To obtain the quantum kernel estimation, the number of quantum measurements per kernel entry was set to 500 (we measured the quantum state in $Z$ basis). This is supported by our noise-model simulations, in which 500 shots were shown to be enough to ensure the quality of the quantum kernel for the objective of this study (see Appendix A). To obtain the quantum kernel matrix for the training data, only the upper triangular entries were computed considering the symmetric nature of the quantum kernel, reducing a computational cost of using the quantum device. In training and testing our QSVC model, we used 20 data points for training and 10 data points (which are separate from the training data) for testing. A total of $105{,}000\ (= 20 \times \frac{21}{2} \times 500)$ quantum measurements were conducted to obtain the quantum kernel and a total of $100{,}000\ (20 \times 10 \times 500)$ shots were conducted to obtain the train-test kernel matrix. In training and testing our QSVR model, we used 40 data points for training and 40 out-of-sample data points for testing. A total of $410{,}000\ (= 40 \times \frac{41}{2} \times 500)$ quantum measurements were conducted to obtain the quantum kernel and a total of $800{,}000\ (40 \times 40 \times 500)$ shots were conducted to obtain the train-test kernel matrix.

To improve the performance of the machine learning models using the quantum kernels, we introduced a scaling hyperparameter $\lambda$ in the quantum feature map (i.e., $x^{(i)} \leftarrow \lambda x^{(i)}$ in the quantum circuit). Such a hyperparameter can calibrate the angles of the rotation gates and affect the quantum feature map in the Hilbert space. The hyperparameter can help improve the performance of the QSVM model (Canatar et al. 2022; Shaydulin and Wild 2022; Suzuki et al. 2023). In the present work, for the classification tasks, the hyperparameter $\lambda$ was set to 1.0, whereas for the regression tasks, $\lambda$ was set to 1.3.

Table 3 Details of quantum-computing experiments using the IonQ Harmony

| Task | Dataset | $\lambda$ | Number of data points | | Number of total shots | |
|---|---|---|---|---|---|---|
| | | | train | test | train | test |
| Classification | | | | | | |
| | Credit card | 1.0 | 20 | 10 | 105,000 | 100,000 |
| | MNIST | 1.0 | 20 | 10 | 105,000 | 100,000 |
| | Fashion-MNIST | 1.0 | 20 | 10 | 105,000 | 100,000 |
| Regression | | | | | | |
| | Financial data | 1.3 | 40 | 40 | 410,000 | 800,000 |
| | Materials data | 1.3 | 40 | 40 | 410,000 | 800,000 |

## 4.2 Preprocessing the materials dataset

We used a dataset for superconducting materials provided by Hamidieh (2018), which was originally compiled by the National Institute of Materials Science in Japan. In this particular dataset, there are 81 features for the critical temperature $T_c$. To reduce the number of features that can be encoded into the NISQ device, the original 81-dimensional vector was reduced into the 4-dimensional vector using principal component analysis (Subasi et al. 2010).

The distribution of critical temperature is concentrated in the low-temperature region. Such a non-normal distribution is not suited for building regression models. To overcome this, we used the Box–Cox transformation (Sakia 1992), which is a statistical technique used to transform a non-normal distribution into a normal distribution. It is often used in regression analysis to improve the performance of the model when the data does not follow a normal distribution. The Box–Cox transformation is defined by the following equation:

$$y^{(\xi)} = \begin{cases} \dfrac{y^\xi - 1}{\xi} & (\xi \neq 0) \\ \log(\xi) & (\xi = 0) \end{cases}.$$

(16)

Here, $y$ is the original data, $y^{(\xi)}$ is the transformed data, and $\xi$ is the Box–Cox transformation parameter. In this study, we used $\xi = 0.15084028$.

## Data Availability

The datasets used in the present study are available at the following:
a) Credit card fraud transaction: https://www.kaggle.com/datasets/mlg-ulb/creditcardfraud
b) MNIST: http://yann.lecun.com/exdb/mnist/index.html
c) Fashion-MNIST: https://github.com/zalandoresearch/fashion-mnist
d) A financial-market dataset used for our QSVR task is tabulated in Supplementary Information (Tables S2 and S3).
e) Superconductor materials: https://archive.ics.uci.edu/ml/datasets/Superconductivty+Data

## Code Availability

A quantum software development kit for reproducing our quantum experiments is available at https://github.com/scsk-quantum/phiqonnect

## Author Contributions

TS conceived the original idea of the work. TM developed the software for the work and conducted the quantum-circuit simulations as well as the quantum experiments. TH conducted the machine learning tasks including the hyperparameter tuning. All the authors analyzed the result data. TS wrote the manuscript. All the authors reviewed the manuscript and commented on it.

## Appendix A: The number of quantum measurements per kernel entry

In Appendix A, we investigate how the number of measurements affects the kernel matrix and the prediction performance (Fig. A1). Three different datasets (the credit card fraudulent transaction dataset, the NMIST dataset, and the Fashion MNIST dataset) were examined. We considered the following conditions for qubit-gate error rates: (i) $p_1 = 0.001$, $p_2 = 0.005$ and (ii) $p_1 = 0.01$, $p_2 = 0.05$. In the case of the lower qubit-gate error rates, the alignment $A(K, K^{\text{noise}})$ remained high and almost unchanged. In the case of the higher qubit-gate error rates, the alignment was slightly improved by increasing the number of measurements. On the other hand, the test accuracy of QSVC models with the higher qubit-gate error rates was very similar to that in the case of the lower error rates. The results suggest that at least 500 shots per kernel entry are enough to ensure the prediction performance of our QSVC models. This is consistent with a recent work by Wang et al. (2021), in which the accuracy is saturated at 500 shots per kernel entry. In the present work, the number of shots per kernel entry was set to 500 for our noisy numerical simulations as well as for quantum experiments using the real quantum device.

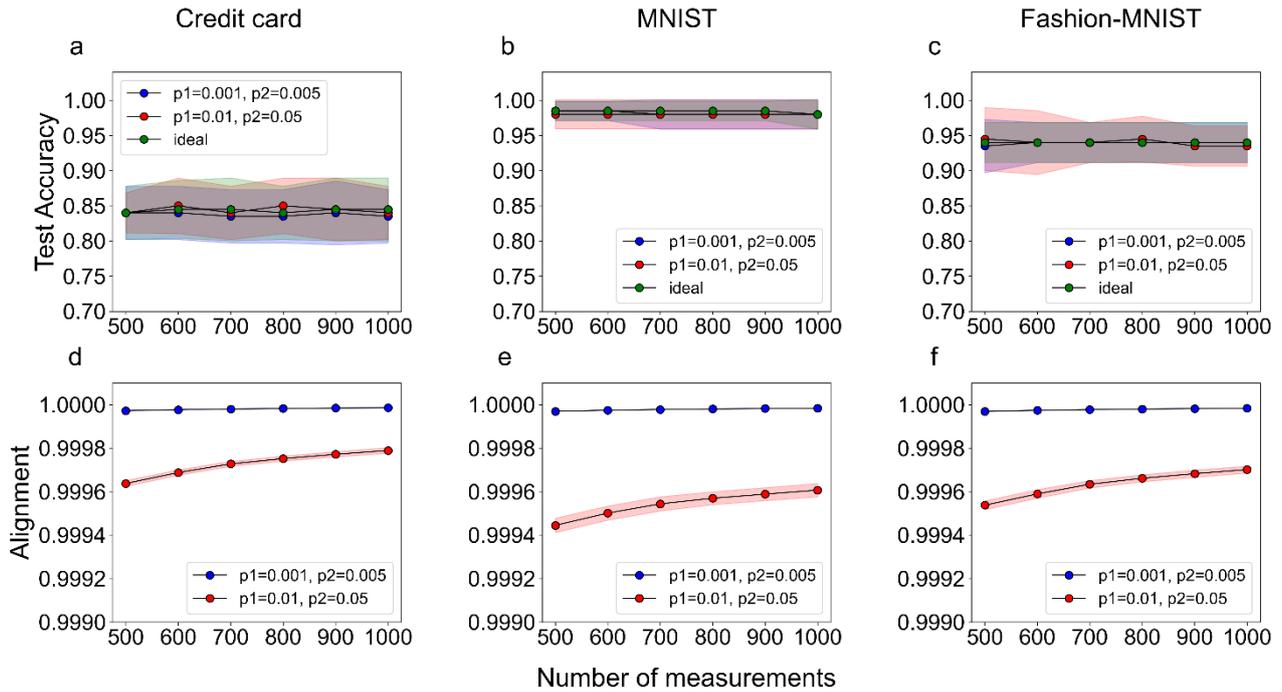

Fig. 8 The effects of shots on the accuracy of QSVM. The error bars indicate the standard deviation obtained from 5 independent seeds. Top panel: test accuracy for (a) credit card dataset, (b) MINST dataset, and (c) Fashion-MNIST dataset. Bottom panel: alignment of the noisy quantum kernel with the noiseless quantum kernel for (d) credit card dataset, (e) MINST dataset, and (f) Fashion-MNIST dataset. In our device-noise-model simulations, we consider the following conditions for single- and two-qubit gate error rates: (i) $p_1 = 0.001$, $p_2 = 0.005$ (indicated by blue); (ii) $p_1 = 0.01$, $p_2 = 0.05$ (indicated by red). Five independent seeds for each dataset were used to obtain the statistical results. The number of training data was 80 and the number of test data was 40.

## Declarations

### Conflict of interest

The authors declare no competing interests.

### Funding

Not applicable.

Supplementary Information for

"Quantum support vector machines for classification and regression on a trapped-ion quantum computer"

Teppei Suzuki[1, *], Takashi Hasebe[1], and Tsubasa Miyazaki[1]

[1] Technology Strategy Division, SCSK Corporation, Toyosu Front, 3-2-20 Toyosu, Koto-ku, Tokyo 135-8110, Japan

**Supplementary Note 1: Comparison between the IonQ Harmony and Aria quantum processors**

Table S1: Quality of the kernel matrix obtained from the IonQ Harmony and Aria quantum processors (4 qubits)

Fig. S1: Comparison in the quantum kernel between the IonQ Harmony and the IonQ Aria (4 qubits).

Fig. S2: The comparison in the quantum kernel between the quantum experiment on the IonQ Aria using 8 qubits and the noiseless simulation.

**The financial dataset used for QSVR**

Table S2: Financial dataset used for QSVR (training data)

Table S3: Financial dataset used for QSVR (test data)

**Supplementary Note 1: Comparison between the IonQ Harmony and Aria quantum processors**

In this section, we report our quantum experiments obtained using two trapped-ion quantum processors released from IonQ, called Harmony and Aria. The Fashion-MNIST dataset was used for the QSVC task in our quantum experiments. There is a significant improvement in terms of qubit gate error rates for the IonQ Aria. In particular, the two-qubit gate error rate is 0.4%, which is a significant improvement from that of the IonQ Harmony (2.7%).

In the first step, we evaluated the quantum kernel estimation derived from the two quantum processors (Table S1). Upon transitioning from the IonQ Harmony to the IonQ Aria, we observed an improvement in the alignment, an increase from 0.993 to 0.998. Consistently, the Frobenius norm between the noiseless kernel and the noisy kernel $\|K - K^{\text{noise}}\|_F$ was significantly reduced.

Table S1: Quality of the kernel matrix obtained from the IonQ Harmony and Aria quantum processors (4 qubits)

|  | Alignment $A(K, K^{\text{noise}})$ | $\|K - K^{\text{noise}}\|_F$ |
|---|---|---|
| IonQ Harmony | 0.993 | 2.436 |
| IonQ Aria | 0.998 | 1.446 |

Next, we took a closer look at the distribution of the deviation in the kernel entry $\Delta K_{ij} = K_{ij}^{\text{noise}} - K_{ij}$ (Fig. S1). About 80% of the kernel matrix elements obtained from the near-term devices were underestimated in comparison with those obtained from the noiseless kernel, owing to the loss of information caused by noise and errors. On the other hand, less than 20% of the kernel matrix elements were somewhat overestimated. The average deviation $\Delta K_{ij}$ in the case of the IonQ Harmony was $-0.091$ whereas that in the case of the IonQ Aria was $-0.051$, indicating that the use of the IonQ Aria leads to a better quality of the quantum kernel. The deviation $\Delta K_{ij}$ obtained from the IonQ Harmony was more broadly distributed. On the other hand, the extent of the underestimation of the kernel entry was significantly reduced when using the IonQ Aria (part c in Fig. S1). Thus, this reduced underestimation played an important role in improving the quality of the quantum kernel. All the results suggest that the IonQ Aria can provide better results in terms of the quality of the quantum kernel matrix.

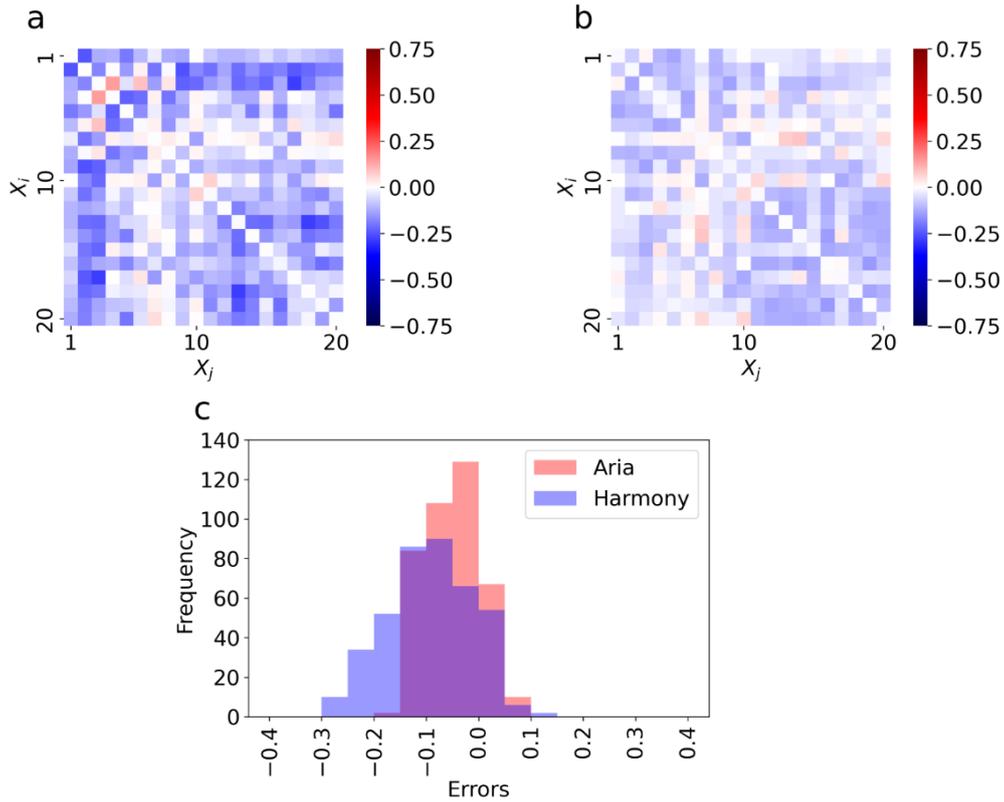

Fig. S1: Comparison in the quantum kernel between the IonQ Harmony and the IonQ Aria (4 qubits). (a) Deviation $\Delta K_{ij}$ obtained using the IonQ Harmony. (b) Deviation $\Delta K_{ij}$ obtained using the IonQ Aria. (c) The histogram for the deviation $\Delta K_{ij}$ for the IonQ Harmony (indicated by blue) and for the IonQ Aria (indicated by red). The Fashion-MNIST dataset was used; 4 qubits were used for both quantum processors.

Motivated by our preliminary results, we further performed our quantum experiment on the IonQ Aria. In this case, we constructed our QSVC models using 8 qubits. The Fashion-MNIST dataset was used for the QSVC task. In our quantum experiment, the alignment between the noiseless kernel and the noisy kernel was 0.977 for the training data, suggesting that the noisy quantum kernel maintained the essential information of the Gram matrix. Our QSVM model achieved 100% accuracy for the test data, without the use of any error mitigation technique (the number of test data was 20). The alignment between the noiseless kernel and the noisy kernel was 0.996, and the average deviation $\Delta K_{ij}$ was $-0.092$ (Fig. S2). The results indicate the usefulness of our quantum circuit on the real device with 8 qubits.

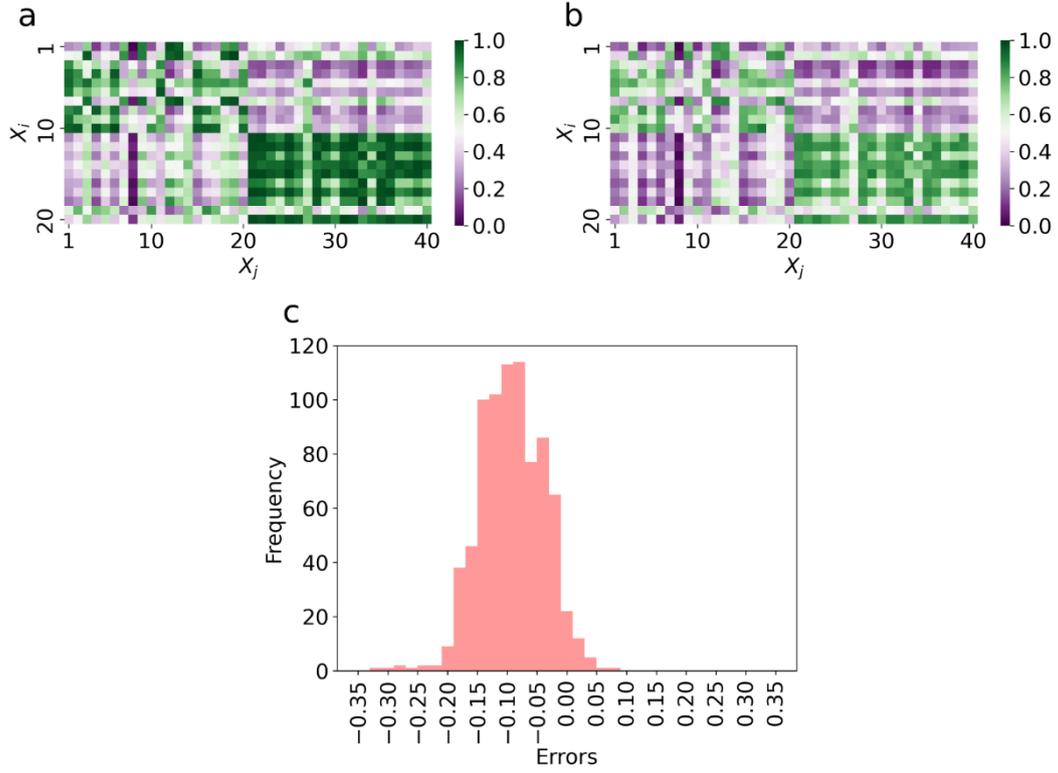

**Fig. S2**: The comparison in the quantum kernel between the quantum experiment on the IonQ Aria using 8 qubits and the noiseless simulation. (a) Train-test quantum kernel for the noiseless simulation. (b) Train-test quantum kernel for the quantum experiment. (c) Histogram for the deviation $\Delta K_{ij}$. We used the Fashion-MNIST dataset (40 data points for training and 20 data points for testing). The alignment of the noisy quantum kernel with the noiseless quantum kernel was 0.996, and the average deviation $\Delta K_{ij}$ was $-0.092$. Using the noisy quantum kernel, our QSVM model achieved 100% accuracy for the test data.

# The financial dataset used for QSVR

Table S2 Financial dataset used for QSVR (training data)

| Number | SSE Index | WTI Crude Oil | US Dollar Index | UK Nickel |
|--------|-----------|---------------|-----------------|-----------|
| 1      | 3676.59   | 68.35         | 92.514          | 19506     |
| 2      | 3536.29   | 74.83         | 94.353          | 18342     |
| 3      | 3581.73   | 69.29         | 92.033          | 19789     |
| 4      | 3272.00   | 96.35         | 106.804         | 21613     |
| 5      | 3547.34   | 83.57         | 94.121          | 19448     |
| 6      | 3228.06   | 97.59         | 107.911         | 19802     |
| 7      | 3607.43   | 66.26         | 96.114          | 20030     |
| 8      | 3387.64   | 108.43        | 104.909         | 21824     |
| 9      | 3067.76   | 108.26        | 103.802         | 30114     |
| 10     | 3253.69   | 112.12        | 98.499          | 31583     |
| 11     | 3662.60   | 70.46         | 92.614          | 19623     |
| 12     | 3489.15   | 92.10         | 96.188          | 24396     |
| 13     | 3047.06   | 104.69        | 103.230         | 31771     |
| 14     | 3567.10   | 68.59         | 92.449          | 19332     |
| 15     | 3693.13   | 68.14         | 92.48           | 20190     |
| 16     | 3579.54   | 78.90         | 95.722          | 20734     |
| 17     | 3562.70   | 69.95         | 96.338          | 20141     |
| 18     | 3186.27   | 94.42         | 106.104         | 22492     |
| 19     | 3429.58   | 91.32         | 95.392          | 23398     |
| 20     | 3607.09   | 72.61         | 92.919          | 19401     |
| 21     | 3518.42   | 82.81         | 93.326          | 19574     |
| 22     | 3613.97   | 71.97         | 93.176          | 19358     |
| 23     | 3609.86   | 83.76         | 93.809          | 20306     |
| 24     | 3593.15   | 82.96         | 93.728          | 20050     |
| 25     | 3275.76   | 97.26         | 106.331         | 21700     |
| 26     | 3409.21   | 111.76        | 104.261         | 23158     |
| 27     | 3555.26   | 82.12         | 94.775          | 22176     |
| 28     | 3597.43   | 82.64         | 94.901          | 22064     |
| 29     | 3284.83   | 120.67        | 104.151         | 27264     |
| 30     | 3613.07   | 73.98         | 93.335          | 19387     |
| 31     | 3625.13   | 71.12         | 96.489          | 19615     |
| 32     | 3558.28   | 81.31         | 93.961          | 19294     |
| 33     | 3189.04   | 88.54         | 105.566         | 22210     |
| 34     | 3167.13   | 94.29         | 99.924          | 32483     |
| 35     | 3516.30   | 68.44         | 92.512          | 19656     |
| 36     | 3582.08   | 76.75         | 96.544          | 20336     |
| 37     | 3524.74   | 69.09         | 93.038          | 19678     |
| 38     | 3522.16   | 68.74         | 92.694          | 19011     |
| 39     | 3576.89   | 65.57         | 96.024          | 19946     |
| 40     | 3637.57   | 72.36         | 95.881          | 20230     |

Table S3 Financial dataset used for QSVR (test data)

| Number | SSE Index | WTI Crude Oil | US Dollar Index | UK Nickel |
| --- | --- | --- | --- | --- |
| 1 | 3561.76 | 80.44 | 94.080 | 18918 |
| 2 | 3582.60 | 83.76 | 93.625 | 19739 |
| 3 | 3569.91 | 85.43 | 95.723 | 22073 |
| 4 | 3477.13 | 65.64 | 92.971 | 18893 |
| 5 | 2928.51 | 98.54 | 101.769 | 32636 |
| 6 | 3675.19 | 69.30 | 92.650 | 19713 |
| 7 | 3186.43 | 114.67 | 101.766 | 28392 |
| 8 | 3498.54 | 80.86 | 93.853 | 19162 |
| 9 | 3642.22 | 73.30 | 93.03 | 19351 |
| 10 | 3146.86 | 110.29 | 102.097 | 27732 |
| 11 | 3054.99 | 106.13 | 104.897 | 27810 |
| 12 | 3398.62 | 105.76 | 104.464 | 22698 |
| 13 | 3238.95 | 121.51 | 103.218 | 28023 |
| 14 | 3546.94 | 80.64 | 94.519 | 18978 |
| 15 | 3225.64 | 106.95 | 100.326 | 33175 |
| 16 | 3675.02 | 72.38 | 96.015 | 19624 |
| 17 | 3582.83 | 75.45 | 93.381 | 18946 |
| 18 | 3462.31 | 95.72 | 96.694 | 24282 |
| 19 | 3130.24 | 115.07 | 101.698 | 28284 |
| 20 | 3320.15 | 104.27 | 104.192 | 24038 |
| 21 | 3004.14 | 103.09 | 103.689 | 28185 |
| 22 | 3271.03 | 114.93 | 98.615 | 32380 |
| 23 | 3446.98 | 66.59 | 93.145 | 19193 |
| 24 | 3282.72 | 99.27 | 98.627 | 33223 |
| 25 | 3186.82 | 104.25 | 99.913 | 32981 |
| 26 | 3715.37 | 70.45 | 92.652 | 19726 |
| 27 | 3267.2 | 106.19 | 103.981 | 24449 |
| 28 | 3573.84 | 66.50 | 96.158 | 19953 |
| 29 | 3451.41 | 91.59 | 96.619 | 24361 |
| 30 | 3465.83 | 93.66 | 95.697 | 23406 |
| 31 | 3628.49 | 72.23 | 93.460 | 19221 |
| 32 | 3586.08 | 79.46 | 96.326 | 20383 |
| 33 | 3544.48 | 84.05 | 93.879 | 19702 |
| 34 | 3263.79 | 122.11 | 102.546 | 28855 |
| 35 | 3266.60 | 107.82 | 97.875 | 32893 |
| 36 | 3656.22 | 72.61 | 92.534 | 20016 |
| 37 | 3567.44 | 81.22 | 95.625 | 21794 |
| 38 | 3595.09 | 72.05 | 96.370 | 20189 |
| 39 | 3313.58 | 104.09 | 107.829 | 21880 |
| 40 | 3214.50 | 105.96 | 99.055 | 32725 |